\documentclass[12pt,english]{article}
\usepackage{lmodern}
\usepackage{lmodern}
\usepackage[T1]{fontenc}
\usepackage{geometry}
\usepackage{color}
\usepackage{babel}
\usepackage{amsmath}
\usepackage{amsthm}
\usepackage{amssymb}
\usepackage{graphicx}

\usepackage[utf8]{inputenc}
\usepackage{float}
\usepackage{xcolor}
\usepackage{tkz-euclide}
\usepackage{caption}
\usepackage{subcaption}
\usepackage{pgfplots}
\usepackage{adjustbox}
\pgfplotsset{compat=1.11}
\usetikzlibrary{arrows,calc}
\tikzset{
>=stealth',
help lines/.style={dashed, thick},
axis/.style={<->},
important line/.style={thick},
connection/.style={thick, dotted},
}

\usepackage[parfill]{parskip}

\usepackage{setspace}
\usepackage[authoryear]{natbib}
\usepackage[unicode=true,
 bookmarks=true,bookmarksnumbered=false,bookmarksopen=false,
 breaklinks=false,pdfborder={0 0 1},backref=false,colorlinks=true]
 {hyperref}

\usepackage{tikz}
\usetikzlibrary{calc}

\makeatletter
	\renewcommand{\abstract}[1]{\def \@abstract {#1}}
	\newcommand{\jelcodes}[1]{\def \@jelcodes {#1}}
	\newcommand{\keywords}[1]{\def \@keywords {#1}}
	\newcommand{\thanknotes}[1]{\def \@thanknotes {#1}}
	\newcommand{\contact}[1]{\def \@contact {#1}}
	\newcommand{\shortauthor}[1]{\def \@shortauthor {#1}}
	\newcommand{\shorttitle}[1]{\def \@shorttitle {#1}}
\makeatother

\jelcodes{}
\thanknotes{}
\shortauthor{}
\shorttitle{}
\abstract{}

\usepackage[hang,flushmargin]{footmisc} 
\usepackage{multibib}
\usepackage{soul}
\usepackage{afterpage}

\makeatletter
\usepackage{fancyhdr}
\makeatother

\newcommand\blfootnote[1]{%
  \begingroup
  \renewcommand\thefootnote{}\footnote{#1}%
  \addtocounter{footnote}{-1}%
  \endgroup
}

\usepackage{authblk}
\makeatletter
\def \maketitle { 
	\thispagestyle{empty}
	\vspace*{0.1in}
	\blfootnote{\textsc{Contact.} \@contact  \@thanknotes\\}
	\begin{center}
	\begin{minipage}{5.2in}
	\begin{center}
	{\large {\textbf{\@title}}}
	
	\vspace{0.2in}
	
	{\textsc{\@author}}
	
	\vspace{0.2in}
	
	{\@date}
	\end{center}
	
	\ifx\@abstract\@empty
	\relax
	\else
	{\small{~~\\ ~~ \\ \textsc{Abstract.} \@abstract}}
	\fi
	
	\ifx\@keywords\@empty
	\relax
	\else
	\vspace{0.2in}
	
	{\small\textsc{Keywords.} \@keywords.}
	\fi
	
	\ifx\@jelcodes\@empty
	\relax
	\else
	{\small\textsc{JEL Codes.} \@jelcodes.}
	\fi
	
	\end{minipage}
	\end{center} }
\makeatother

\theoremstyle{definition}

\theoremstyle{definition}
 
\theoremstyle{plain}

\theoremstyle{plain}

\theoremstyle{definition}

\theoremstyle{remark}

\theoremstyle{plain}

\theoremstyle{plain}

\theoremstyle{remark}

\theoremstyle{plain}

\usepackage{color}
\usepackage{babel}

\hypersetup{
  colorlinks,
  linkcolor={blue!80!black},
  citecolor={blue!50!black},
  urlcolor={red!50!black},
}

\makeatother

\providecommand{\assumptionname}{Assumption}
\providecommand{\claimname}{Claim}
\providecommand{\conditionname}{Condition}
\providecommand{\corollaryname}{Corollary}
\providecommand{\definitionname}{Definition}
\providecommand{\examplename}{Example}
\providecommand{\lemmaname}{Lemma}
\providecommand{\notationname}{Notation}
\providecommand{\propositionname}{Proposition}
\providecommand{\theoremname}{Theorem}

\title{\Large Familiarity Facilitates Adoption \\ \normalsize Evidence from Electric Vehicles} 
\author{\textsc{Jonathan Libgober ~~ Ruozi Song}}

\affil{University of Southern California}

\abstract{This paper shows that a non-price intervention which increased the prevalence of a new technology facilitated its further adoption. The BlueLA program put Electric Vehicles (EVs) for public use in many heavily trafficked areas, primarily (but not exclusively) aimed at low-to-middle income households. We show, using data on subsidies for these households and a difference-in differences strategy, that BlueLA is associated with a 33\% increase of new EV adoptions, justifying a substantial portion of public investment. While the program provides a substitute to car ownership, our findings are consistent with the hypothesis that increasing familiarity with EVs could facilitate adoption.}


\shortauthor{Libgober and Song}
\shorttitle{BlueLA}

\contact{libgober@usc.edu, ruozison@usc.edu. We thank Matt Kahn, Rob Metcalfe, Emily Nix, Paulina Oliva, Geert Ridder and Jeff Weaver for helpful comments,  and Natasha Denisyuk for excellent research assistance.  We are also particularly grateful for input from the California Air Resources Board (CARB) and LADOT on this project, especially Sam Gregor, John Massetti, Meri Miles, Anthony Poggi, Anita Tang, and Collin Weigel.}

\begin{document}

\maketitle

\section{Introduction}

Whenever a new and more socially efficient technology emerges, society has a natural interest in spurring its adoption. The traditional approach to achieving this end has been to directly subsidize its purchase. If an individual does not internalize their positive externality, subsidies can align private incentives with the social optimum. 

However, if consumers are insufficiently familiar with the new technology---something one would expect to usually be the case---this logic can run into problems. Computing the optimal subsidy requires determining the difference between an individual's privately optimal adoption incentive and the social optimum. But when adoption requires individuals to abandon a known alternative, uncertainty over the former may prevent adoption even with an added monetary benefit. For subsidies to have an impact, an even greater amount may be needed to compensate for this uncertainty, potentially making them prohibitively costly to be effective at all. The fact that consumers, by their nature, have more experience with older technologies, provides a powerful default in their favor.

This paper studies how non-price interventions that increasing familiarity\footnote{In this paper, we use the term ``familiarity'' in an informal sense, as a catch-all to refer to the gap between a consumer's true (objective) surplus from consumption and the believed (subjective) surplus from consumption. We do not distinguish, for instance, risk aversion on the benefits of EV adoption versus awareness per se.} can spur adoption, focusing on Electric Vehicles (EVs) adoption.  Plainly, increasing adoptions of EVs is a major public policy goal, in the United States at many levels of government, as well as internationally. The larger aim is to reduce dependence on gas vehicles and increasing energy efficiency toward creating a more sustainable economy. Toward this end, the Infrastructure Investment and Jobs Act passed in late 2021 included billions of dollars for EV Charging stations and other investments in electric vehicles. California in particular has devoted considerable attention to increasing EV usage; in 2020, Governor Newsom signed an executive order requiring that in 2035, all cars and passenger trucks sold in California to be zero-emission vehicles. Despite this urgency, the takeup of EVs has been relatively slow. While the rate of takeup has been increasing, with a 12.9\% increase in EV sales in Q2 2022 from Q2 2021, this still only amounted to 5.6 \% of the total market \citep{AutoNews}. There appears to be substantial work to do to take EV adoption rates to a point where achieving the 100\% target be practically feasible and non-disruptive.

We analyze the impact of the BlueLA car sharing program on EV adoption. This program placed Electric Vehicles throughout low-to-middle income neighborhoods of Los Angeles in busy areas, with the idea of increasing the accessibility of green technology to communities that traditionally benefit less from them. For our purposes, the most important features of the BlueLA program were the following: First, \emph{anyone} had the ability to use an EV through the program, often in the immediate area. Second, even when not using an EV, the location of the program means that EVs were placed in desirable central locations, obtainable due to its status as a public program. Third, along the lines of it being a public program, usage costs were quite low, typically cheaper than competitive car sharing programs.

Despite being a direct substitute to car ownership, we find that the BlueLA program has so far had a substantial \emph{positive} impact on EV adoption. We interpret this finding as suggesting increasing familiarity with EVs may be an important component of the push to spur adoption.\footnote{\citet{Garziano2015solarpanels} documents a similar phenomenon in the context of solar panels, namely that increased visibility of solar panels in a neighborhood leads to more adoption.} To benchmark our results, we compare these findings to other interventions. Even our most conservative estimates find that the impact of BlueLA was comparable to the impact of a direct 10\% subsidy, and some rough quantification of the value of this increase in takeup in itself would justify a large part of the State's investment.\footnote{From our conversations with The State of California's agency in charge of environmental programs, CARB, this possibility was not a direct consideration. CARB invested \$4.6 Million into the program; the overall cost was \$31 million.} We note that our data used comes from those individuals who used this particular subsidy for the purpose of EV adoption; thus, our preferred interpretation is that the subsidy was \emph{twice as effective} in areas which were exposed to the BlueLA treatment. It is also worth bearing in mind that the increased adoption is only a secondary benefit of the program, and one that is (arguably) more significant only at the state level.\footnote{For instance, the company running the program profits when more people use it, and thus have an incentive to discourage car ownership. Likewise, local entities would arguably benefit more from easy access to amenities.}

We document a number of other patterns which are suggestive of our mechanism that an informational story is responsible for the finding. First, when analyzing the trend over time, we see that the effect takes approximately one year to show up. This is consistent with lags in information diffusion. Second, we consider the impact of charger availability, treating this as a proxy for the true average utility for EV adoption in an area (e.g., \citet{li2016compatibility}). We find that this does not interact with any of our results, suggesting that variation in latent utility across zip codes is not influenced by the program. Third, we find that effects are localized---we document no effect if we look at zip codes adjacent to those with BlueLA adoption, rather than those directly effected. This suggests interaction with the program at the local level is important, consistent with information diffusion being facilitated by immediate visibility. Fourth, we find no brand specific effect---which, given that all cars in the program are the same make and model, might be plausible a priori---suggesting our findings are not driven by a ``taste-based'' channel, which may influence the latent utility from consumption outside of information. Taken together, the patterns we document as well as our understanding of the mechanisms of the program suggest an informational mechanism as responsible.

We therefore advance evidence that non-price instruments may be important for increasing takeup, and that policymakers interested in increasing EV adoption should consider this as an important channel. The EV market features both high costs to implement adoption as well as apparent susceptibility of purchasing behavior to information prevision. On costs, \cite{muehlegger2022policy} survey a number of findings and document that even conservatively, the elasticity estimates on EV adoption suggest that the cost of inducing a new EV adoption to be larger than the face value of the subsidy, if not larger than a direct purchase of EVs themselves. On consumer awareness, several studies have found many consumers do not know the subsidies available or are appear mistaken about basic characteristics of EVs \citep{jinslowik2017},  or appear to misoptimize when evaluating the costs and benefits of EV ownership \citep{bushnelletal2022}. Furthermore, market pressures do not appear conducive to new product adoption on their own. For instance, \cite{cahill2014new}, \cite{de2018dismissive}, \cite{lynes2018dealerships} show that dealers tend to discourage EV adoption, often to the point of being actively deceptive (with Tesla being a major exception). In other words, dealers display an apparent preference for gas cars, and thus consumers who are not otherwise already inclined to purchase EVs might not naturally be pushed toward doing so. Thus, we hope our findings underscore that initiatives which facilitate greater consumer familiarity with EVs may lead to a higher degree of subsidy effectiveness, and potentially correct the apparent market failure associated with the frictions toward their adoption.

\section{Background} 

\subsection{The BlueLA Program}
In this section, we provide background information about the BlueLA program, and elaborate on certain institutional details which will play important roles in motivating and interpreting our analysis.

\begin{figure}[H]
\centering
\includegraphics[scale=.065]{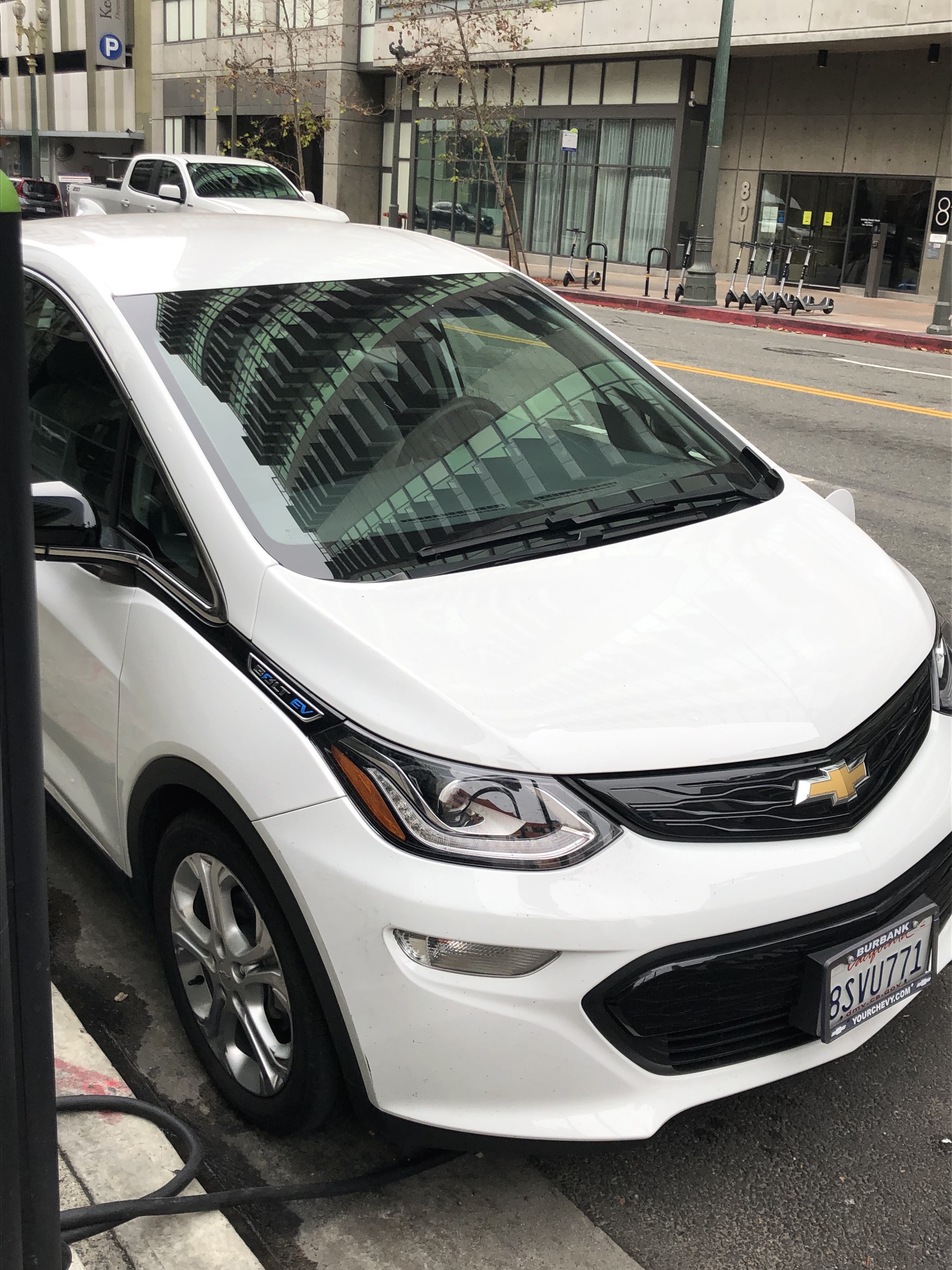}
\includegraphics[scale=.065]{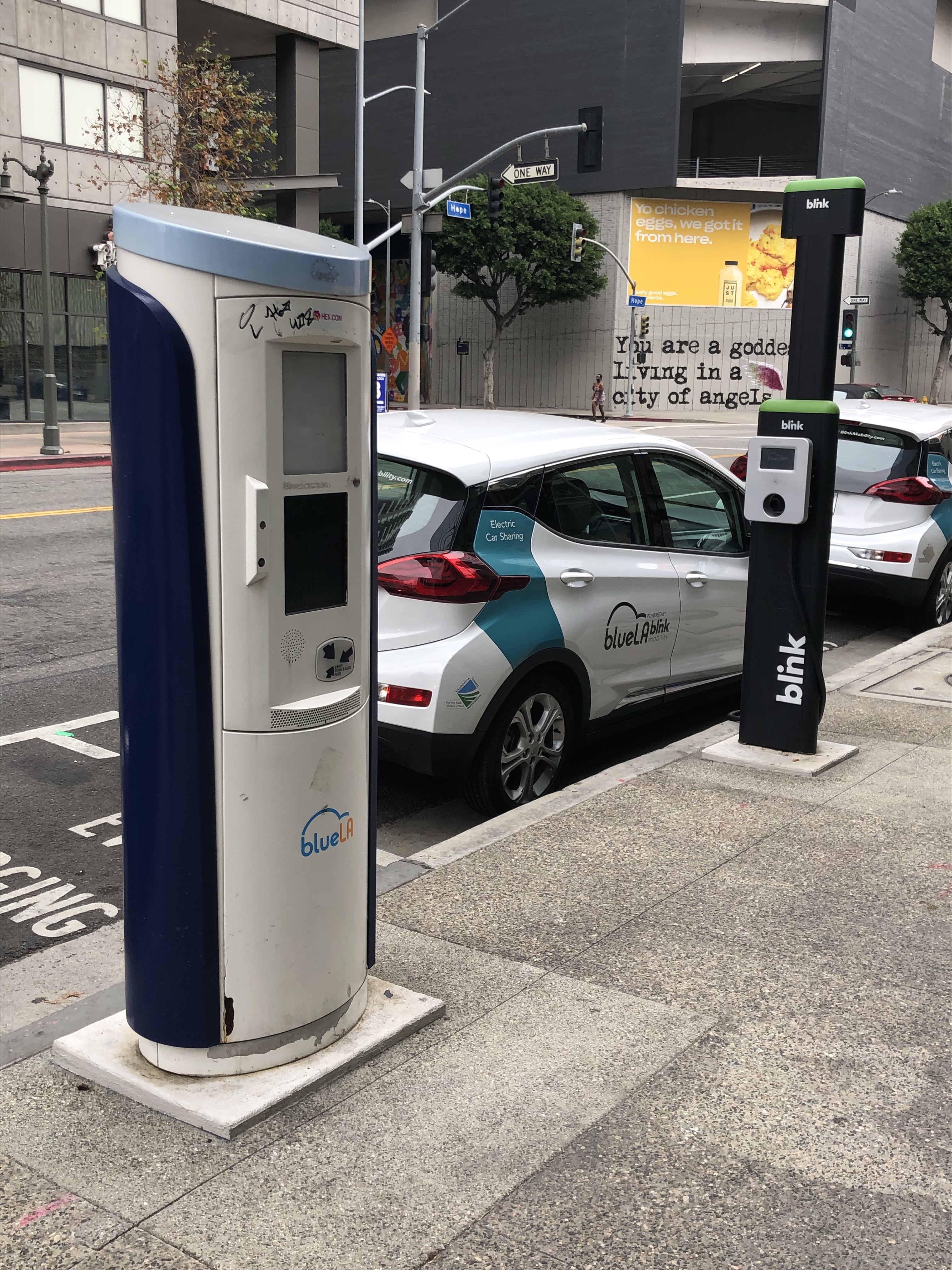} \\ 
\includegraphics[scale=.065]{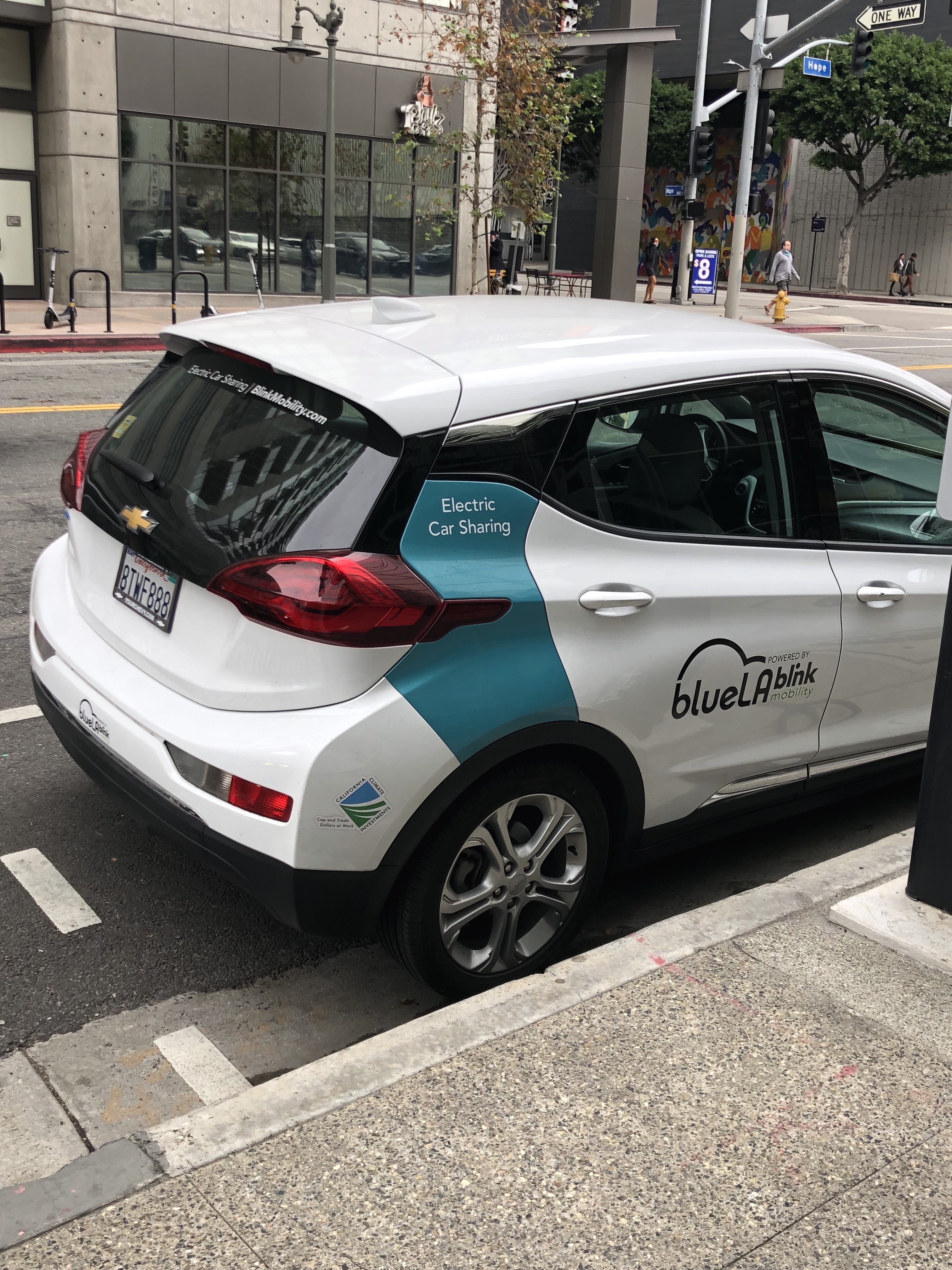}
\includegraphics[scale=.065]{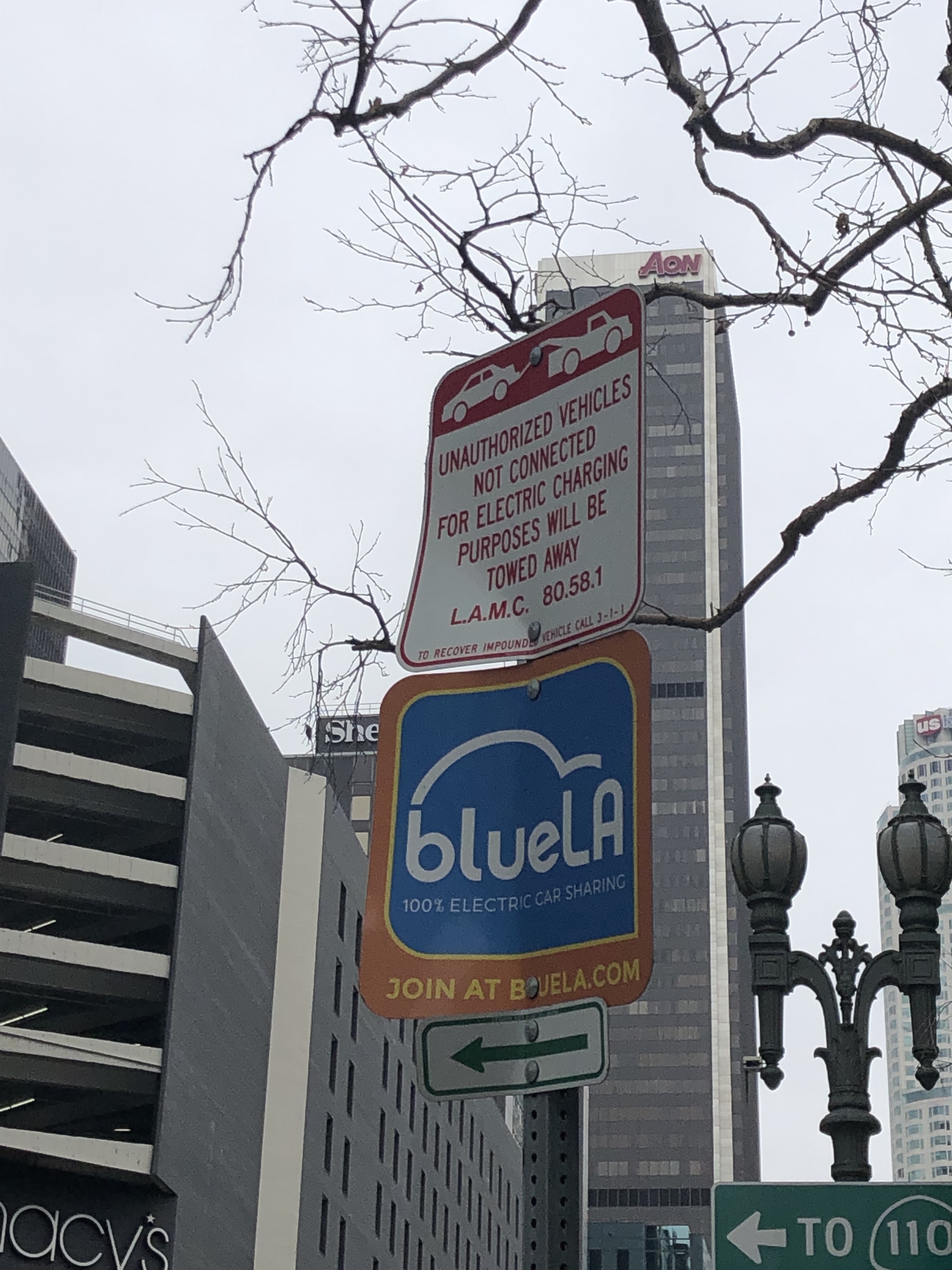}
\caption{A Representative BlueLA Charging Station}  \label{fig:cars}
\end{figure}

BlueLA is a car sharing program which is particularly aimed at increasing accessibility of green technology to income groups which are traditionally excluded from it. These cars can be found at charging stations located in various neighborhoods in LA, but in keeping with the goal of the program, typically in areas where they would be used by low-income residents. Figure \ref{fig:cars} shows a typical BlueLA charging station, which also shows the standard design of a BlueLA car. For our purposes, the notable feature is that these cars are highly visible and in areas in which they are easily accessible. The program itself is a public-private partnership, operated by a private company (Blink Mobility, as of September 2020, and the French company Bollor\'e prior to that) and funded by the California Air Resources Board (CARB). The LA Department of Transportation (LADOT), in addition to providing some funding, also provides infrastructure for the program; for instance, providing permits to locate car chargers in heavily trafficked areas which would otherwise be used for street parking. However, the cars themselves and the chargers that are used are both purchased and maintained by Blink. All cars are highly standardized, following the same designs and same type, specifically Chevy Bolts.\footnote{Chevy Bolts have been the exclusive car model used since Blink has operated the program; Bollor\'e used their own vehicle, the Bollor\'e Bluecar.} CARB's initial grant to BlueLA was for \$4.6 million dollars, while the program had around 18 million dollars of private investment and additional grants for a grand total cost of  \$31 Million.

\begin{figure}[H]
\centering
\includegraphics[scale=.65]{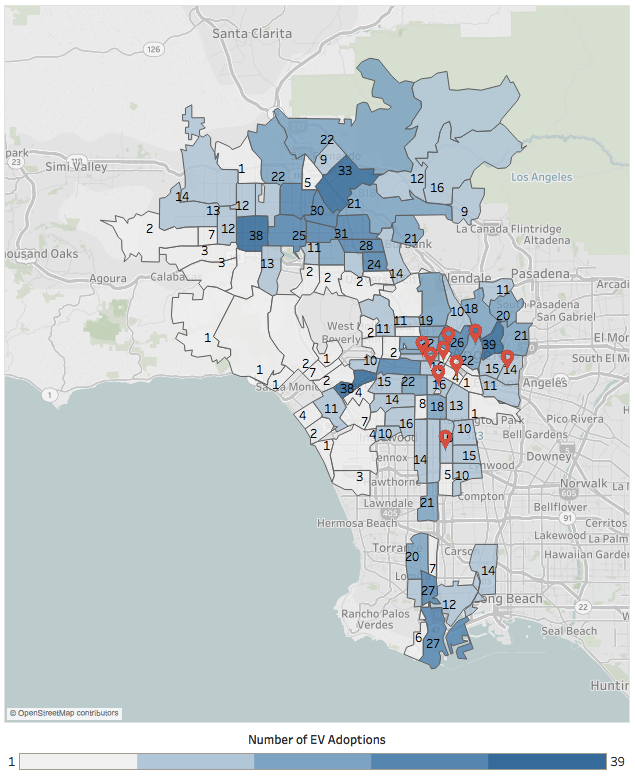} \\ 
\caption{EV Adoption and BlueLA Stations}  \label{fig:map}
\footnotesize{\setstretch{0.5} \begin{flushleft} Note: This map displays the numbers of accumulated EV adoption by zipcodes in the City of Los Angeles from 2015 to 2021, and the locations of BlueLA stations by the end of 2021.\end{flushleft}}
\end{figure}

\begin{figure}[H]
\centering
\includegraphics[scale=.75]{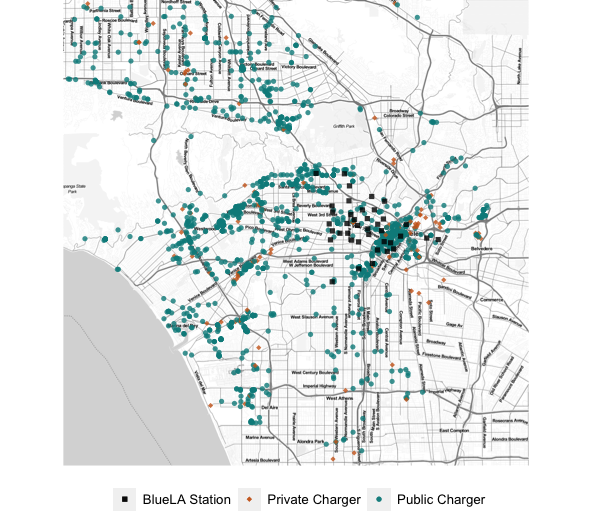} \\ 
\caption{Public EV Chargers in Los Angeles}  \label{fig:map_charger}
\footnotesize{\setstretch{0.5} \begin{flushleft} Note: Locations of public and private EV chargers, as of December 2021, are from the Department of Energy's Alternative Fuels Data Center. Black dots represent BlueLA stations set up by the end of 2021. \end{flushleft}}
\end{figure}

Each BlueLA charging station usually involves enough chargers for a small number of cars, typically around five. Importantly, the chargers are themselves only usable exclusively by cars that are part of the BlueLA program, and the spots are designated exclusively for these cars as well. There are a number of implications of this fact; perhaps most obviously, the chargers which are installed for the program cannot possibly be used by other EV cars that are available. But it also is often a key source of the difficulty in introducing BlueLA more widely, in that finding a location where a charger can be placed is typically one main challenge administratively. BlueLA has formal criteria which they use to help determine and evaluate spot locations; these include density, proximity to transit (with priority to locations near hubs), employment density (with priority to employment/retail centers), income levels (with priority to high density areas with affordable housing/multi-family housing), transit modal shares, and suitability of EVs. However, in our conversations with BlueLA, it was conveyed to us that the main stumbling block was in finding parking spaces for the cars and chargers. Figure \ref{fig:map} shows the location of zipcodes with BlueLA Chargers at the end of 2021, as well as the total number of accumulated adoptions, which forms the basis of our empirical analysis; Figure \ref{fig:map_charger} shows the map of the locations of EV Chargers as of March 2022.    
 
\begin{figure}[H]
\centering
\includegraphics[scale=.5]{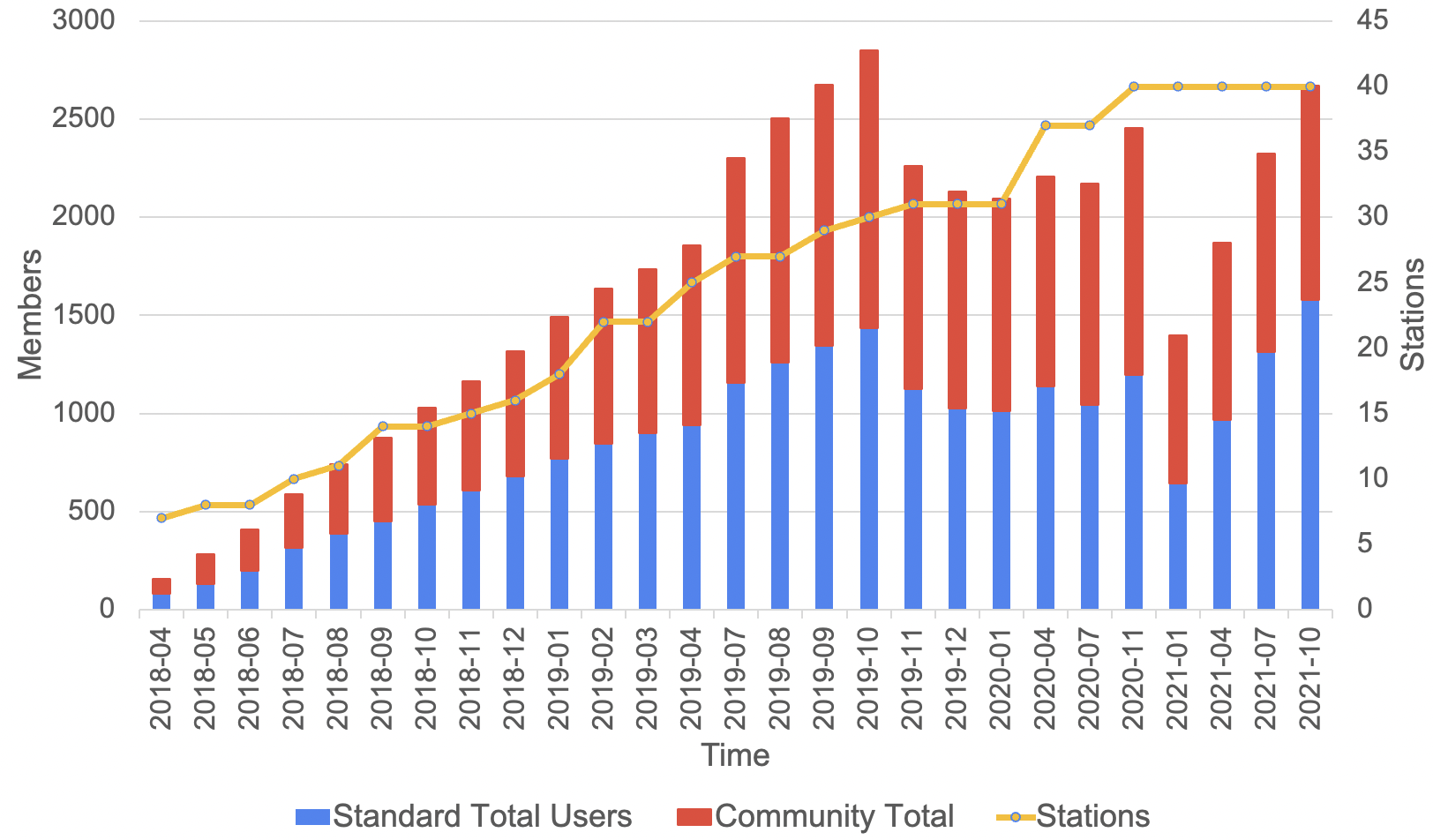} \\ 
\includegraphics[scale=.5]{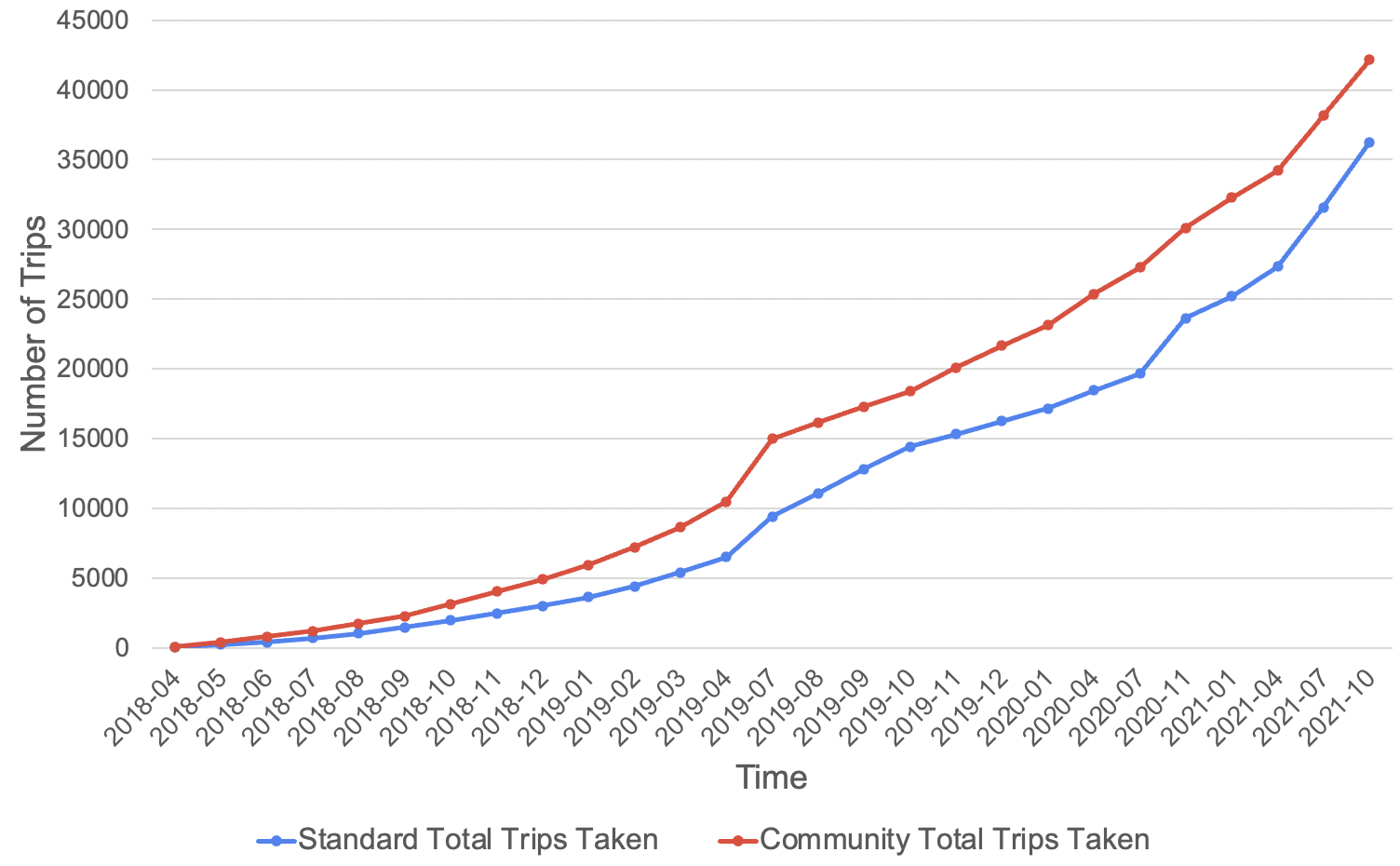} \\ 
\caption{Members, Trips Taken and Stations of BlueLA program. Source: BlueLA, Report to City of Los Angeles}  \label{fig:bluela_detail}
\end{figure}

BlueLA started to be operable in April 2018, and has had a steady rate of growth since its start. In April 2019, the program had provided 80 electric vehicles, 130 charge points, and 26 charging stations, having  served nearly 2000 members and over 12,000 trips. Since then the program has continued to grow; as of with currently over 300 cars and 40 locations and expansion plans continue. Since its start, there have been three different class of users; the standard membership costs 5 dollars per month and 20 cents per minute of use; community members (who qualify based on income) pay 1 dollar per month and 15 cents per minute of use; and one-month trial members, who pay 40 cents per minute but do not pay any up front fee.\footnote{For reference, a current Zipcar membership costs 7 dollars per month and 11 dollars per hour, although often in private lots, in contrast to BlueLA chargers which take street parking spots.} Packages are also offered for extended rentals of either 3 hours or 5 hours at lower rates. Community members have taken about 54 \% of all trips. Figure \ref{fig:bluela_detail} shows the number of members and trips taken broken down by membership type.

Initially the agenda of the program was completely unrelated to EV Adoption, simply to provide carsharing in underserved areas using green technology. However, around 2020 the BlueLA program explicitly sought to educate the community about EVs. We note that advertising around BlueLA often highlights the benefits of driving Electric Cars. For instance, on the Blink Mobility blog, while certain aspects of the program are emphasized, significant attention is also paid to the convenience of owning Electric Vehicles. In fact, one of their blogposts, entitled \emph{Thinking about buying an EV? Rent One by the Hour!}, advances our proposed mechanism explicitly, suggesting users could use the program as an intermediate step toward deciding whether they wanted to purchase an EV---while also (naturally) arguing that users should also consider the program as a primary option \emph{instead} of owning a car. The latter force, if significant, would push EV adoption downward, and for this reason we interpret our estimates as lower bounds to the effect of increased familiarity. 

\subsection{Data} 
To estimate the effect of the BlueLA rollout, we combine two data sources; first, we obtained  rebate data from the Clean Cars 4 All Program (CC4A). This rebate data is publicly available data at the transaction level on the usage of a particular subsidy for EV usage targeted at individuals from lower income brackets. This data includes the subsidy, the vehicle purchased, and the zipcode in which the recipient of the subsidy levels. Second, we obtained from LADOT the documented opening date for each location. We use this to determine when the program enters each particular zip code. 

We note that for all relevant observations, whenever BlueLA has ``entered'' a zipcode, it has remained in that zipcode throughout the rest of our sample. Since the location of the subsidy usage is at the zipcode level, we aggregate each specific location to be at the zipcode level as well. This allows us to merge the two data sets so that the unit of observations is the number of purchases of EVs in a given zipcode.  

For some additional robustness checks, we supplement this with data on the number of charging stations in each zipcode. This data is from the Alternative Fuels Data Center, which has also been used in other papers (e.g., \citet{li2016compatibility}). This provides us the number of public chargers present in any zipcode for each quarter. 

Table \ref{tab:sum_cohort} presents some summary statistics regarding the zip codes into which BlueLA has been introduced. Anticipating some of our methodology to come, we group zip codes into ``cohorts'' depending on when the first BlueLA charging station was introduced. As we explain more below, part of our analysis will estimate treatment effects for each cohort.\footnote{In this table, data on income comes from 2016-2020 American Community Survey, and are 5-Year estimates, and data on EV adoptions comes from \href{https://data.ca.gov/dataset/vehicle-fuel-type-count-by-zip-code}{the State of California}.} We can therefore use this data to see how our estimates vary with zip code characteristics, which will allow us to assess mechanisms which may be suggestive of our findings.  

\section{Identification Strategy}\label{sect:strategy}

We follow two strategies toward determining the impact of BlueLA on adoption. First, and perhaps most straightforwardly, we estimate the following equation: 

\begin{equation} 
NewEVs_{i,t}= \alpha_{i}BlueLA_{i} + \beta BlueLA_{i} \times Entry_{i,t}+  \delta X_{i,t}+ \gamma_{i} +\eta_{t} + \varepsilon_{it}, \label{eq:main}
\end{equation} 

\noindent where $NewEVs_{i,t}$ denotes the number of new EV adoptions in zipcode $i$ at time $t$, $BlueLA_{i}$ is an indicator variable that denotes the event that a BlueLA charging station \emph{ever} exists in zipcode $i$, whereas $Entry_{i,t}$ is an indicator variable that denotes the event that a BlueLA charging station exists in zipcode $i$ at time $t$. Letting $\beta_{i,t}$ denotes the the number of adoptions at time $t$ in zipcode $i$ due to the introduction of BlueLA, then in the above equation, 
\begin{equation}
\beta = \sum_{i,t} w_{i,t} \beta_{i,t}. \label{eq:weightsexp}
\end{equation} Under the assumption that the $w_{i,t}$ are all positive and normalized to sum to 1, $\beta$ represents a weighted average of each $\beta_{i,t}$ coefficient. However, a substantial amount of recent work has shown that the positivity assumption need not hold; see \citet{roth2022s} as well as \citet{de2022survey} for recent surveys of this literature. We address this issue after presenting our results.

\begin{figure}[H]
\centering
\begin{subfigure}{0.8\textwidth}
\centering
\includegraphics[width=0.9\linewidth]{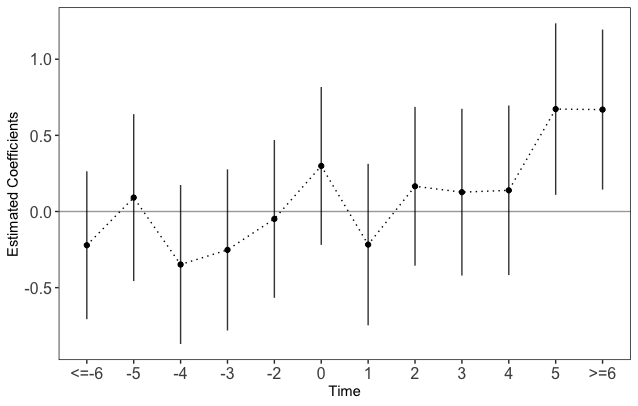}
\caption{\footnotesize{Without Charger Control}} 
\label{fig:coef_plot_nocharger}
\end{subfigure}
\smallskip
\begin{subfigure}{0.8\textwidth}
\centering
\includegraphics[width=0.9\linewidth]{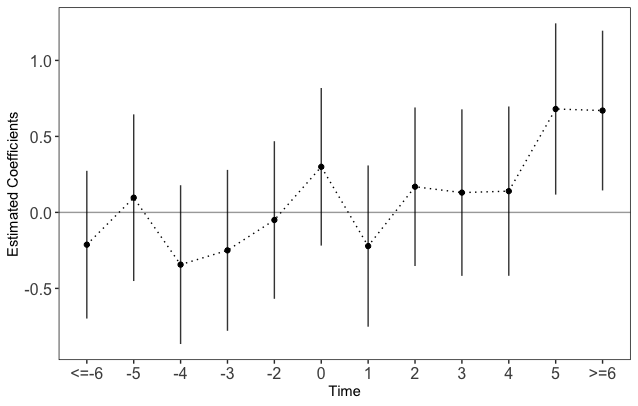}
\caption{\footnotesize{With Charger Control}} 
\label{fig:coef_plot_charger}
\end{subfigure}
\caption{Event study for introduction of BlueLA on new vehicle adoptions} 
\footnotesize{\setstretch{0.5} \begin{flushleft} Note: The figure reports the coefficient plot of linear panel regressions with two-way fixed effects. Standard errors are two-way cluster standard errors at year-quarter and zipcode level at 5\% significance level. \end{flushleft}}
\end{figure}

Our identification assumption for $\beta$ is that the trend of EV adoption in neighborhoods with BlueLA car sharing stations would have been the same as other neighborhoods had BlueLA never entered. We test this assumption by following an event study approach using the following equation: 
\begin{equation}
N_{it}=\sum_{k \ge -6, k \ne -1}^{k=6}\delta_kD_{it}^k +\gamma_i+\eta_t+\epsilon_{it}, \label{pre-trend}
\end{equation}
where the dummy variables $D_{it}^k$ indicate time compared to the event. We define $s_i$ as the date when zipcode $i$ had the first BlueLA station; $D_{it}^{-6}=1$ if $t-s_i \le -6$ (quarters) and 0 otherwise; $D_{it}^k=1$ if $t-s_i = k$ and 0 otherwise for k=-4,-3,...,5; and $D_{it}^{6}=1$ if $t-s_i \ge 6$ and 0 otherwise. Note that the dummy for $k = -1$ is omitted in equation (\ref{pre-trend}) so that the post-treatment effects are relative to the period 5 quarters prior to the establishment of BlueLA stations. We also consider an event study where we include $X_{it}$ as a right hand side variable; the $\delta_{k}$ coefficients from Equation (\ref{pre-trend}) are plotted in Figure \ref{fig:coef_plot_nocharger}, whereas the coefficients where chargers are included as a control are plotted in Figure \ref{fig:coef_plot_charger}. Overall, these look quite similar to one another, and do not feature any statistically significant coefficients prior to time 0.\footnote{We include both of these plots since, in general, having a statistically significant coefficient without controls does not necessarily imply that these coefficients will remain significant when controls are added.} In fact, as the coefficient $\xi$ tends to be close to 0, we do not comment on this further.

One concern with conducting pre-trend test using event study framework with Two-Way Fixed Effect is that the test can be under-powered, passing a pre-trend test can at best be uninformative and at worst introduce additional bias into a design. The issues involved are discussed in \cite{roth2019pre},  which also provides methodology for testing whether the pre-trend tests have sufficient power. To verify the robustness of pre-trend test that uses a Two-Way Fixed Effect framework, we first check the maximum linear pre-trend our test can detect, given a power of 0.8 \footnote{A power of 0.8 means that the pre-trend test can significantly identify pre-trend correctly 80\% of times.}. Then we calculate what the expected value of the coefficients would have been conditional on passing the pre-trend test had this in fact been the true trend.

\begin{figure}[H]
\centering
\begin{subfigure}{0.8\textwidth}
\centering
\includegraphics[width=0.9\linewidth]{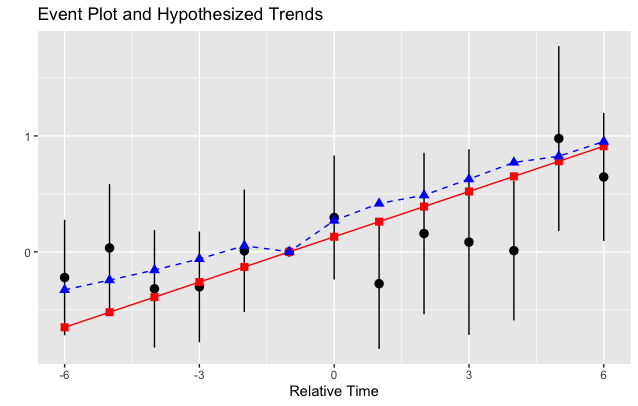}
\caption{\footnotesize{Without Charger Control}} 
\end{subfigure}
\smallskip
\begin{subfigure}{0.8\textwidth}
\centering
\includegraphics[width=0.9\linewidth]{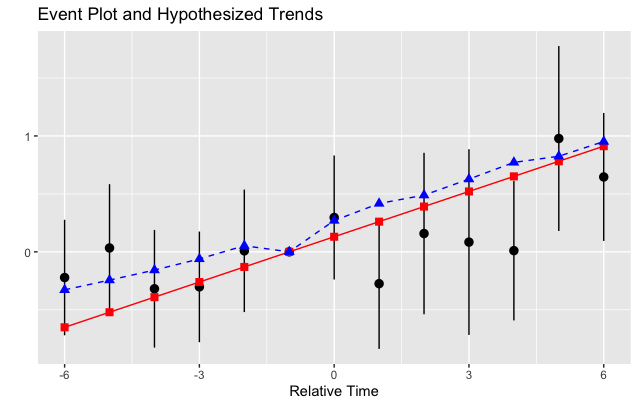}
\caption{\footnotesize{With Charger Control}} 
\end{subfigure} 
\caption{Static Event Study Analysis Testing for Pre-Trend}  \label{paralleltrendplot}
\footnotesize{\setstretch{0.5} \begin{flushleft} These two figures report the parallel trend test, proposed in \cite{roth2019pre}, without chargers and conditional on chargers respectively. The black dots represent the estimated coefficients from the conventional Two-Way Fixed Effect model. Standard errors are two-way cluster standard errors at year-quarter and zipcode level at 5\% significance level. The red squares indicate the maximum linear trend this conventional Two-Way Fixed Effect pre-trend test can detect, given a power of 0.8. The blue triangles are the expectations of coefficients conditional on passing the parallel pre-trend. \end{flushleft}}
\end{figure}

Figure \ref{paralleltrendplot} represents the results from our pre-trend tests and subsequent power analyses with and without controlling number of public chargers. First, plotting each of the point estimates as well as the confidence intervals for each one, we note that none of the coefficients on $\delta_{k}$ are statistically significant from 0 for $k < 0$, normalizing $\delta_{-1}=0$. It also shows that the calculation as described in the previous paragraph suggests our tests have sufficient power, since the maximum detectable linear trend can match the trend of treatment effect well. Thus, it is unlikely that the estimated effect of BlueLA stations is driven by some underlying pre-trend that we fail to detect, as the pre-trend test should have enough power to pick up such pre-trend. Also, the tests with and without charger control are very similar, which suggests that the number of public chargers shouldn't bias our estimates. 

Of independent interest is that the statistical significance emerges roughly one year after the introduction of the BlueLA chargers. This is consistent with our proposed mechanism, since information diffusion tends to take time as consumers need to have the opportunity to notice the introduction of the program, and also make a car purchasing decision. As we would not expect this to be immediate, it seems sensible to observe a delay in the impact of the program on EV adoption. 

Our second empirical strategy, which complements the direct estimation of Equation (\ref{eq:main})
is to implement the Difference-in-Difference (DiD) estimator of \cite{callaway2021difference}. This estimator avoids the issues of conventional two-way fixed effects regressions with heterogeneous treatment effect, in particular issues related to negative weights. Instead of pooling time-varying treatments and calculating the average treatment effect directly, treatment cohorts are defined by the groups of zipcodes that share the same starting date of first BlueLA station. Control units are those zipcodes where no BlueLA station ever set up. Then, the average treatment effect can be estimated for each treatment cohort by comparing each treatment cohort and all the control units, which reduces to the canonical DiD setup with two groups and two periods. These cohort-specific estimates are additionally useful in our discussion of mechanisms, as we analyze the extent to which variation in the cohorts correlates with the resulting estimates of the treatment effect.

\section{Results} 

\subsection{Estimating the BlueLA Impact} 

We now present our main regression results on the purchases of new EVs via the introduction of the BlueLA program. We presented results from a linear panel model in Table \ref{DID}; since our dataset involves a large number of 0 purchases, we also estimate a negative binomial model and present the results in Table \ref{DIDNB}. We note that, while zipcodes having a BlueLA charging station by itself does not predict new EV adoptions, we find statistically significant increases in adoption once the chargers are installed. It is important to keep in mind that the coefficients on each variable correspond to a number of new adoptions, at the zipcode/quarter level. Overall, EV adoption rates are somewhat low over the sample, and so to interpret these magnitudes it is perhaps helpful to consider the \emph{total} number of EV adoptions implied by these estimates, as shown in Table \ref{BOEC}. Our sample involves 112 zipcodes, which are the set of zipcodes which have had at least one adoption in our sample; keeping in mind that these are at the quarterly level, we therefore have multiplying by 448 gives a total estimate for the number of EV adoptions attributable to the variable of interest. On the other hand, the total number of EV Adoptions in LA in 2020 in these zipcodes is 304. Therefore, we should not expect any of these coefficients to be particularly large. However, translating the estimates by adjusting the scale accordingly, Table \ref{DID} suggests BlueLA lead to, most conservatively, 99 new adoptions over a year period, and the estimate from Table \ref{DIDNB} suggests 136 new adoptions. 

To help interpret the magnitude of these coefficients, we can compare to the impacts of other interventions, most notably direct subsidy interventions. A number of studies estimating the impact of a subsidy have been conducted; \cite{muehlegger2018subsidizing} estimate that a 10\% subsidy on EVs for low- and middle-income individuals corresponds to 32\%-34\% increase in new EVs. Using the most conservative estimate from Table \ref{DID} gives a similar effect---specifically, 32.6\%---of the BlueLA program; noting that the negative binomial model is interpreted as a percentage, the same calculation suggests an even larger increase of 44.7\%.\footnote{The negative binomial regression identifies the percent change in EV purchases; we then convert it to average increase in annual EV adoption by taking the product of the percentage increase and the average annual EV adoption. We then divide by the total number of adoptions in 2020 to obtain this percentage.}   We note that the subsidy studied in \cite{muehlegger2018subsidizing} amounts to roughly 5000 per car; using this number as a conversion between how to have developed a similar effect via subsidy, this would correspond to a benefit of about \$495000 per year using most conservative Table \ref{DID} estimate and about \$680000 per year using the most conservative Table \ref{DIDNB} estimate. While relatively small, this appears sizable for a side effect of the program (whereas the sole benefit of a subsidy is through the channel of car ownership). Of course, it is also worth commenting that these individuals have access to the subsidies as well, so in some sense this might be ``double counting'' the benefits to an EV purchase. On the other hand, while it is also worth being cautious with such extrapolations, these rough calculations suggest that in one year, at least 10\% of CARB's initial investment in BlueLA could have been recovered through this channel; in fact, the benefit we identify is larger since this amounts to the average effect taken over the entire six year sample, and could indeed be durable in the future. Thus, our results suggest that a non-trivial portion of the initial investment of CARB\footnote{Our focus on CARB is due to the fact that it was the entity whose primary interest was in reducing dependence of EVs, as per the mandate from state government; the other organizations involved, like LADOT or Blink Mobility, arguably have other more pressing motivates like assisting local transport or profit from the program.} was paid back for through the channel we identify. 

\subsubsection{Testing the Interpretation of $\beta$} 

As mentioned in Section \ref{sect:strategy}, recent research has noted that staggered rollout designs can suffer from interpretability issues and bias, when the treatment effect is heterogeneous across groups and periods so that the weights in equation (\ref{eq:weightsexp}) are not all positive. While the regressions highlighted in the previous section are the most straightforward, a concern with interpretability and bias would emerge if in fact our estimates were subject to the same issues. We now follow a number of strategies to address these issues. 

\begin{figure}[H]
\centering
\includegraphics[width=0.8\textwidth]{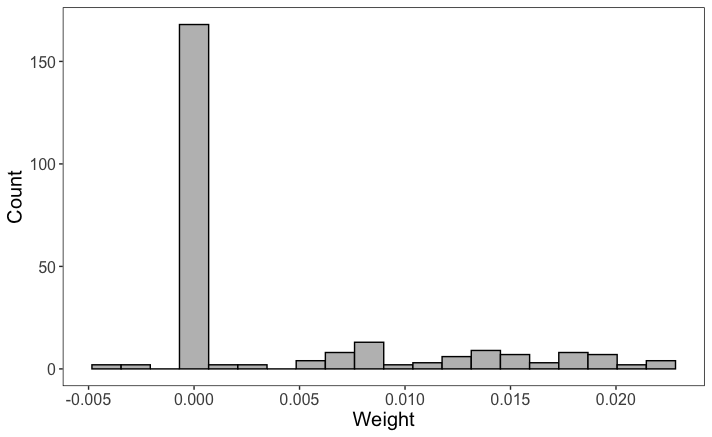}
\caption{Histogram of Two-Way Fixed Effect Weights}  \label{nagative_weight_test}
\end{figure}

The first is to follow \cite{de2020two} to test whether the negative weight issue occurs. Figure \ref{nagative_weight_test} shows the actual weights of each group-period pair in Two-Way Fixed Effect regression (\ref{eq:weightsexp}). Only 4 out of 252 pairs suffer from negative weights, and even then these are very small. So, while perhaps possible, it seems implausible that such estimates would invalidate our results, at worst tempering the previous conclusions somewhat. 

\begin{figure}[H]
\centering
\includegraphics[width=0.6\linewidth]{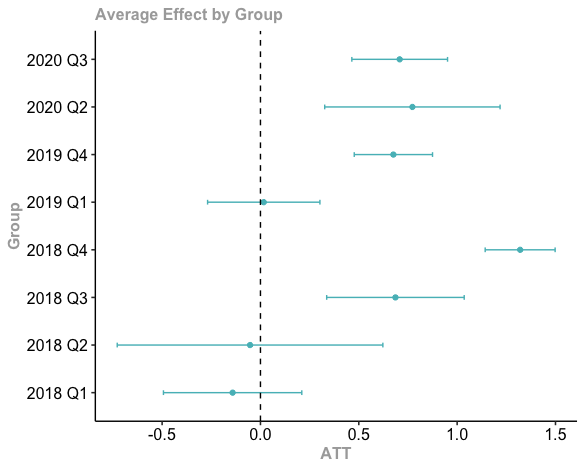}
\caption{Average Effects of BlueLA Station on EV Adoption by Cohort} \label{csdid}
\end{figure}

The second is to implement the \cite{callaway2021difference} estimator, described above in Section \ref{sect:strategy}, noting that this estimator is not subject to these particular concerns. This produces treatment effect estimates by cohort, where (i) cohorts are defined by the groups of zipcodes that share the same starting date of first BlueLA station, and (ii) control units are those zipcodes where no BlueLA station ever set up. Figure \ref{csdid} shows the average treatment effect of the BlueLA program by cohort. Our study period covers 2015 to 2021 at quarter and zipcode level. The first BlueLA stations were up and running in April 2018, 13 quarters after. Therefore, for instance, Group 15 includes all the zipcodes where the BlueLA program started since July 2018 (15 quarters after Jan 2015). Aside from being less susceptible to bias related to treatment staggering, this figure also illustrates more precisely which specific cohorts have had a more significant effect on takeup. Interestingly, we find positive average treatment effects for all but three cohorts (2018Q1, 2018Q2, and 2019Q1), which we cannot distinguish from 0. Other than these three cohorts, the effect seems to be fairly stable. Table \ref{tab:sum_cohort} shows the summary statistics by cohorts in terms of average household income, number of public chargers, and number of universal EV takeup. We study correlations between these summary statistics and the treatment effects derived in Section \ref{sect:drivers}.

To translate these cohort-specific estimates into an aggregated treatment effect comparable to those from our other panel models, we take an weighted average over the treatment effects of different cohorts estimated via this approach, putting more weight with larger group sizes, as shown in Table \ref{BOEC}. The average treatment effect we obtain correspond to a coefficient of $\beta$ equal to 0.391; an even \emph{more} significant impact of the program an adoption, suggesting that the potential sources of bias related to the staggering of treatments actually leads to an understatement of the program's effectiveness. This estimate corresponds to a total yearly increase of 175 new EV adoptions, which using the same reasoning as before, would correspond to 57.6\% increase.

\subsection{Interpreting the Results Using Institutional Details}

We summarize some key points regarding the institutional details that are important for understanding these findings. 

First, it is worth keeping in mind that the BlueLA program is \emph{a direct substitute} to car ownership. It is entirely possible that some individuals would have needed to purchase a car but no longer need to due to the program; in fact, the program advocates this, as one would expect. Therefore, insofar as we are estimating the impact of an increase in familiarity due to adoption, this particular program could in principle by itself decrease adoption. In other words, while our proposed channel is through the increase in awareness and availability of EVs, our intervention should have the direct effect of lowering EV takeup.   

Second, the introduction of a BlueLA chargers do not appear endogenous to any decisions that would influence purchasing. The main variable of concern to us is charger availability; the reason being, it could be that BlueLA chooses locations based on EV charger availability, and that EV charger availability further lowers the cost of owning an EV. Controlling for this does not influence our conclusion. 

One could imagine that changing neighborhood characteristics from gentrification both predict the introduction of BlueLA as well as EV purchasing behavior; however, our data comes from entirely low income members, who are also disproportionately the targets of the BlueLA program (as increased community engagement with green technology is an explicit mandate). In other words, since our sample is already restricted based on income, it would likely not be possible for changing demographics of a particular zipcode to be correlated with both the introduction of BlueLA as well as new EV adoption.

\section{Mechanisms} 

We now present direct evidence that the informational channel is most significant. We seek to disentangle informational drivers (e.g., awareness or understanding) from those which influence the latent utility from consumption. Above, we showed that chargers did not influence the relative impact of the program. We now present a number of other findings which underscore this.  

\subsection{Drivers of the Treatment Effect}  \label{sect:drivers}
\begin{figure}[H]
\centering
\includegraphics[scale=.5]{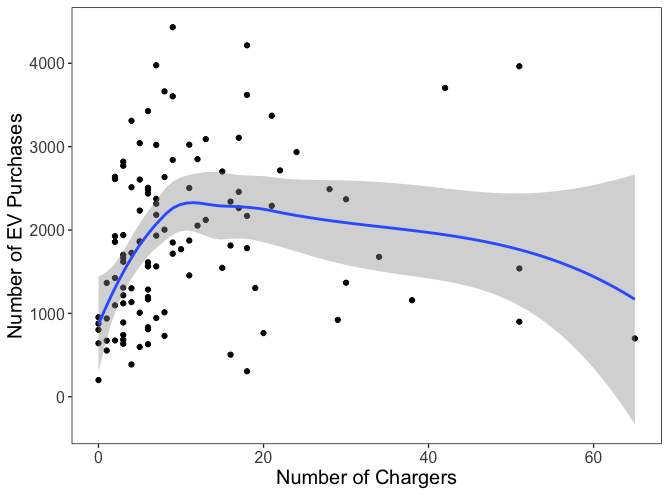} \\ 
\caption{Public EV Chargers and Overall EV takeups in Los Angeles}  \label{fig:plot_adoption_charger}
\footnotesize{\setstretch{0.5} \begin{flushleft} Note: This figure reports the relationship between the number of public EV chargers and overall EV takeups across zipcodes in Los Angeles by January 2021. Public charger information is from the Department of Energy's Alternative Fuels Data Center and we calculate the number of total public chargers available for each zipcode by Jan 2021. The number of EV takeups is from \href{https://data.ca.gov/dataset/vehicle-fuel-type-count-by-zip-code}{the State of California}, which provides the universal EV counts by zipcodes by Jan 2021 .  \end{flushleft}}
\end{figure}

An advantage of having access to cohort-specific estimates is to use heterogeneity in these cohorts to obtain a rough understanding of which properties drives the effects we document. Toward this end, we compute simple correlations between the summary statistics reported in Table \ref{tab:sum_cohort} and the estimated treatment effects at the cohort levels. To be clear, we do not seek to be overly formal in this analysis, but view this as simply suggestive. We find the correlation between income and the treatment effect is 0.527; such a correlation appears sensible, as areas with higher income likely contain more individuals who would consider buying a new car. Perhaps surprisingly, we find a \emph{negative} correlation between the number of chargers and our treatment effect, -.357; though one might expect this to be positive as well (since living in an area with more chargers suggests more convenience for EV usage), this is consistent with our story since areas with lots of chargers likely have a higher degree of ``baseline awareness.'' This is consistent with number of chargers itself also spurring an increase in adoption, something we document in Figure \ref{fig:plot_adoption_charger}. Still, insofar as chargers represent a measure of ``average consumption utility from EV usage,'' it suggests our results are not driven by a direct change to this variable caused by the introduction of BlueLA.

\subsection{Spillovers Across zipcodes}

One question of interest is whether the effect we identify is isolated to the areas where the EVs are located. Specifically, there are various reasons why the informational channel we identify might have an effect; one channel is usage, but another channel is akin to advertising, specifically seeing more EVs in the local area. While the former channel is likely only to be relevant in close proximity to the charging stations, the latter need not be. Therefore, we look at whether there is any impact on zipcodes adjacent to those with EV chargers. Our results are negative on this point;  We also follow the same two-way fixed effect design as in section \ref{sect:strategy}, Equation (\ref{pre-trend}) to estimate the effect of BlueLA stations in adjacent zipcodes on EV adoption. Table \ref{tab:twfe_adj} shows that there is no significant spillover effect. This suggests that the channel is more local, either requiring a greater degree of contact or actual easy usage in order for a significant effect to show up.

\subsection{Car Model Influence} 
\begin{figure}[H]
\centering
\includegraphics[width=0.8\linewidth]{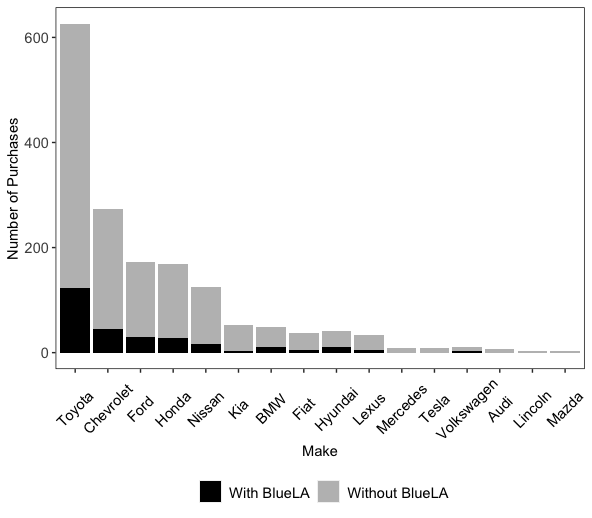}
\caption{Top 10 Makes of Replacement Vehicles in CC4A} \label{fig:car_make}
\end{figure}

An alternative explanation of EV purchase increase can be that consumers like the driving experience of the specific car make used in BlueLA program, which has nothing to do with learning about EVs in general. BlueLA only provides Chevy Bolt EVs. If this is the case, then we should observe that consumers tend to purchase Chevy Bolt EV as the replacement vehicle. However, Figure \ref{fig:car_make} shows that the most popular car make is Toyota among participants in the CC4A rebate program, rather than Chevy Bolt.

\section{Conclusion}

This paper documents that zipcodes for which BlueLA charging stations were introduced experienced a greater increase in EV adoption relative to other zipcodes which did not experience any such introduction. Given that the program functionally makes EVs easily accessible for all individuals in an a given area, our interpretation of this finding is that greater familiarity and access to EVs can increase their adoption, perhaps overcoming a signifcant hurdle toward the goal of universal adoption advanced by several policymakers. 

On the one hand, the obvious caveats to this conclusion of course apply. We view our analysis as underscoring the message that it would likely be worthwhile for policymakers and economists to think about non-price instruments that could help facilitate an increase in adoption, particularly as it relates to information, familiarity or accessibility. On the other hand, based on existing work on the car market as well as green technology as a whole, this gap appears to remain a significant once facing consumers. Hence another area for future work may be in analyzing why markets fail to correct such gaps in the first place, as well as what kinds of mechanisms may be introduced to combat them. 
 
\bibliography{EVsBib}
\bibliographystyle{ecta}

\appendix

\newpage
\section{Tables} 

\begin{table}[H]
\centering
\begin{tabular}{cccc}
\hline \hline
\textbf{Cohort} & \textbf{Avg. Income (\$)} & \textbf{Avg. \# Chargers} & \textbf{Total Number of EVs} \\
\hline
2018 Q1              & 48,322                & 6                     & 1,970                         \\
2018 Q2              & 40,481                & 27                        & 1,104                         \\
2018 Q3              & 44,083                & 11                        & 2,038                         \\
2018 Q4              & 49,068                & 4                         & 1,726                         \\
2019 Q1              & 37,216                & 54                        & 2,202                         \\
2019 Q4              & 56,389                & 12                        & 3,023                         \\
2020 Q2              & 58,195                & 7                       & 2,120                         \\
2020 Q3              & 45,499                & 52                        & 1,539  \\
\hline \hline
\end{tabular}
\caption{Summary Statistics by Groups of zipcodes in Jan 2021} \label{tab:sum_cohort}
\end{table}
\footnotesize{\setstretch{0.5} \begin{flushleft} Note: A cohort is defined as a group of zipcodes that share the same starting date as the first BlueLA stations. The cohort name indicates the quarter when the first BlueLA stations were set up. Income data is from the 2016-2020 American Community Survey 5-Year estimates. Public charger information is from the Department of Energy's Alternative Fuels Data Center and we calculate the number of total public chargers available for each zip code by Jan 2021. The total number of EVs is from \href{https://data.ca.gov/dataset/vehicle-fuel-type-count-by-zip-code}{the State of California}, which provides the universal EV counts by zipcodes by Jan 2021. \end{flushleft}}

\newpage
\begin{table}[H] 
\centering 
\begin{adjustbox}{width=0.5\columnwidth,center}
\begin{tabular}{@{\extracolsep{5pt}}lccc} 
\\[-1.8ex]\hline 
\hline \\[-1.8ex] 
 & \multicolumn{3}{c}{\textit{Dependent variable:}} \\ 
\cline{2-4} 
\\[-1.8ex] & \multicolumn{3}{c}{NewEVs} \\ 
\\[-1.8ex] & (1) & (2) & (3)\\ 
\hline \\[-1.8ex] 
 BlueLA & 0.071 & 0.076 &  \\ 
  & (0.074) & (0.074) &  \\ 
  & & & \\ 
 N\_charger & $-$0.004 & $-$0.002 & 0.002 \\ 
  & (0.003) & (0.003) & (0.002) \\ 
  & & & \\ 
 BlueLA $\times$ Entry & 0.250$^{**}$ & 0.231$^{**}$ & 0.221$^{**}$ \\ 
  & (0.110) & (0.110) & (0.102) \\ 
\hline \\[-1.8ex] 
Year FEs & Yes &  &  \\ 
Quarter FEs & Yes &  &  \\ 
Year-Quarter FEs &  & Yes & Yes \\ 
Zipcode FEs &  &  & Yes \\ 
Observations & 3,136 & 3,136 & 3,136 \\ 
R$^{2}$ & 0.056 & 0.106 & 0.282 \\ 
Adjusted R$^{2}$ & 0.052 & 0.098 & 0.249 \\ 
\hline 
\hline \\[-1.8ex] 
\end{tabular} 
\end{adjustbox}
 \caption{Linear Panel Regressions: Main Results on the Influence of BlueLA Charging Stations on Adoption of EVs}
  \label{DID} 
\footnotesize{\setstretch{0.5} \begin{flushleft} Note: This table reports the effect of BlueLA Stations on the number of incremental EV purchases, using linear panel regressions. The unit of observation is at year-quarter and zip code levels. BlueLA indicates whether there are or there will be BlueLA stations in a zip code. Entry indicates whether there are already BlueLA stations in a zip code. The number of public chargers available is included as a control. Standard errors are two-way cluster standard errors at year-quarter and zip code levels. $^{*}$p$<$0.1; $^{**}$p$<$0.05; $^{***}$p$<$0.01\end{flushleft}}
\end{table}

\begin{table}[H] 
\centering 
\begin{adjustbox}{width=0.8\columnwidth,center}
\begin{tabular}{@{\extracolsep{5pt}}lccccccc} 
\\[-1.8ex]\hline 
\hline \\[-1.8ex] 
 & \multicolumn{7}{c}{\textit{Dependent variable: }\text{NewEVs}} \\ 
\cline{2-8} 
\\[-1.8ex] & \multicolumn{3}{c}{Negative Binomial} & & \multicolumn{3}{c}{Poisson} \\ 
\\[-1.8ex] & (1) & (2) & (3) & &  (4) & (5) & (6)\\ 
\hline \\[-1.8ex] 
BlueLA & 0.115 & 0.107 & & & 0.028 & 0.033 & \\ 
  & (0.216) & (0.214) & & & (0.228) & (0.228) &  \\ 
  & & & & & & &\\ 
 N\_charger & $-$0.004 & $-$0.005 & $-$0.005 & & -0.002 & -0.003  & -0.003 \\ 
  & (0.005) & (0.005) & (0.005) & & (0.004) & (0.004) & (0.004)\\ 
  & & & & & & &\\ 
 BlueLA $\times$ Entry & 0.587$^{***}$ & 0.577$^{***}$ & 0.600$^{***}$ & & 0.580$^{***}$ & 0.572$^{***}$ & 0.577$^{***}$\\ 
  & (0.145) & (0.146) & (0.140) & & (0.133) & (0.133) & (0.128) \\ 
& & & & & & &\\
\hline \\[-1.8ex] 
Year FEs & Yes &  & &  & Yes &  &  \\ 
Quarter FEs & Yes &  & & & Yes &  &  \\ 
Year-Quarter FEs & & Yes & Yes & & & Yes & Yes \\ 
Zipcode FEs & &  & Yes &  & &  & Yes \\ 
Observations & 3,136 & 3,136 & 3,136 & & 3,136 & 3,136 & 3,136 \\ 
Log-Likelihood & -2847.759 & -2856.311 & -2856.437 & & -2878.434  & -2891.864 & -2891.875  \\
\hline 
\hline \\[-1.8ex] 
\end{tabular} 
\end{adjustbox}
 \caption{Negative Binomial and Poisson Panel Regressions: Main Results on the Influence of BlueLA Charging Stations on Adoption of EVs}
  \label{DIDNB} 
\footnotesize{\setstretch{0.5} \begin{flushleft} Note: This table reports the effect of BlueLA Stations on the number of incremental EV purchases, using negative binomial panel regressions. The unit of observation is at year-quarter and zip code levels. BlueLA indicates whether there are or there will be BlueLA stations in a zip code. Entry indicates whether there are already BlueLA stations in a zip code. The number of public chargers available is included as a control. $^{*}$p$<$0.1; $^{**}$p$<$0.05; $^{***}$p$<$0.01\end{flushleft}}
\end{table} 

\newpage
\begin{table}[H] 
\centering 
\begin{adjustbox}{width=0.7\columnwidth,center}
\begin{tabular}{@{\extracolsep{5pt}}lccc} 
\\[-1.8ex]\hline 
\hline \\[-1.8ex] 
\\[-1.8ex] & TWFE & NB & CSDID\\ 
\hline \\[-1.8ex] 
Coefficient & 0.221$^{**}$ & 0.600$^{***}$ & 0.391$^{**}$ \\ 
 & (0.102) & (0.140) & (0.181)\\
 Induced annual EV adoption & 99 & 136 & 175 \\ 
 Equivalent percentage increase in 2020 & 32.6\% & 44.7\% & 57.6\% \\
\hline 
\hline \\[-1.8ex] 
\end{tabular}
\end{adjustbox}
 \caption{Comparison Between Specifications: Main Results on the Influence of BlueLA Charging Stations on Adoption of EVs}
  \label{BOEC} 
\footnotesize{\setstretch{0.5} \begin{flushleft} Note: This table compares the effect of BlueLA Stations on the adoption of EVs using linear panel regression with two-way fixed effects in Table \ref{DID}, negative binomial panel regression in Table \ref{DIDNB}, and the average effect from Callaway and Sant'Anna's estimator in Figure \ref{csdid}. $^{*}$p$<$0.1; $^{**}$p$<$0.05; $^{***}$p$<$0.01\end{flushleft}}
\end{table}  

\newpage
\begin{table}[H] 
\centering 
\begin{adjustbox}{width=0.6\columnwidth,center}
\begin{tabular}{@{\extracolsep{5pt}}lccc} 
\\[-1.8ex]\hline 
\hline \\[-1.8ex] 
 & \multicolumn{3}{c}{\textit{Dependent variable:}} \\ 
\cline{2-4} 
\\[-1.8ex] & \multicolumn{3}{c}{NewEVs} \\ 
\\[-1.8ex] & (1) & (2) & (3)\\ 
\hline \\[-1.8ex] 
  BlueLA\_adjacent & $-$0.041 & $-$0.038 &  \\ 
  & (0.073) & (0.074) &  \\ 
  & & & \\ 
 N\_charger & $-$0.005 & $-$0.003 & 0.003 \\ 
  & (0.004) & (0.004) & (0.002) \\ 
  & & & \\ 
 BlueLA\_adjacent $\times$ Entry\_adjacent & 0.094 & 0.086 & 0.109 \\ 
  & (0.105) & (0.105) & (0.099) \\ 
  & & & \\ 
\hline \\[-1.8ex] 
Year FEs & Yes &  & \\ 
Quarter FEs & Yes &  &  \\ 
Year-Quarter FEs &  & Yes & Yes \\ 
Zipcode FEs &  &  & Yes \\ 
Observations & 2,716 & 2,716 & 2,716 \\ 
R$^{2}$ & 0.044 & 0.093 & 0.284 \\ 
Adjusted R$^{2}$ & 0.040 & 0.083 & 0.250 \\ 
\hline 
\hline \\[-1.8ex] 
\end{tabular} 
\end{adjustbox}
 \caption{The Effect of BlueLA Stations in Adjacent Zipcodes: Two-Way Fixed Effects}
  \label{tab:twfe_adj} 
\footnotesize{\setstretch{0.5} \begin{flushleft} Note: This table reports the effect of BlueLA stations in an adjacent zipcode on the number of incremental EV purchases in zip codes without any BlueLA stations, using a two-way fixed effect regression. Zipcodes (15 out of 112) with at least one BlueLA station during our study period are excluded from the sample. Of the remaining zip codes, 24 of them are adjacent to zip codes with at least one BlueLA station. The unit of observation is at year-quarter and zip code levels. BlueLA\_adjacent indicates whether there are or there will be BlueLA stations in an adjacent zip code. Entry\_adjacent indicates whether there are already BlueLA stations in an adjacent zip code. The number of public chargers available is included as a control. Standard errors are two-way cluster standard errors at year-quarter and zip code levels. $^{*}$p$<$0.1; $^{**}$p$<$0.05; $^{***}$p$<$0.01\end{flushleft}}
\end{table}

\newpage
\section{Figures}

\newpage

\newpage

\newpage

\newpage

\newpage 

\newpage

\begin{figure}[H]
\centering
\includegraphics[width=0.6\linewidth]{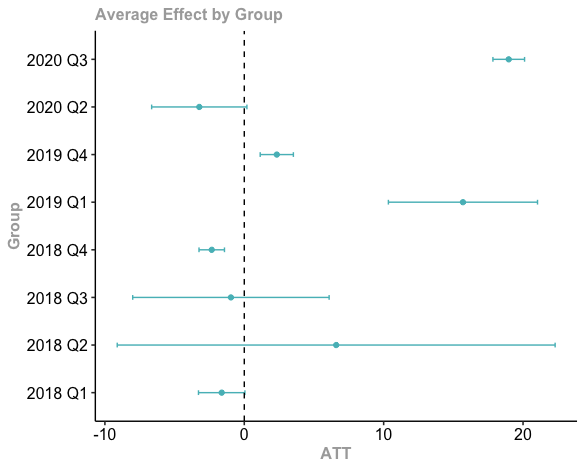}
\caption{Average Effects of BlueLA Station on Number of Chargers by Cohort} \label{csdid_charger}
\footnotesize{\setstretch{0.5} \begin{flushleft} Group is defined as the group of zipcodes that share the same starting date of its first BlueLA station.  \end{flushleft}}
\end{figure}

\newpage

\begin{figure}[H]
\centering
\begin{subfigure}{0.8\textwidth}
\centering
\includegraphics[width=0.9\linewidth]{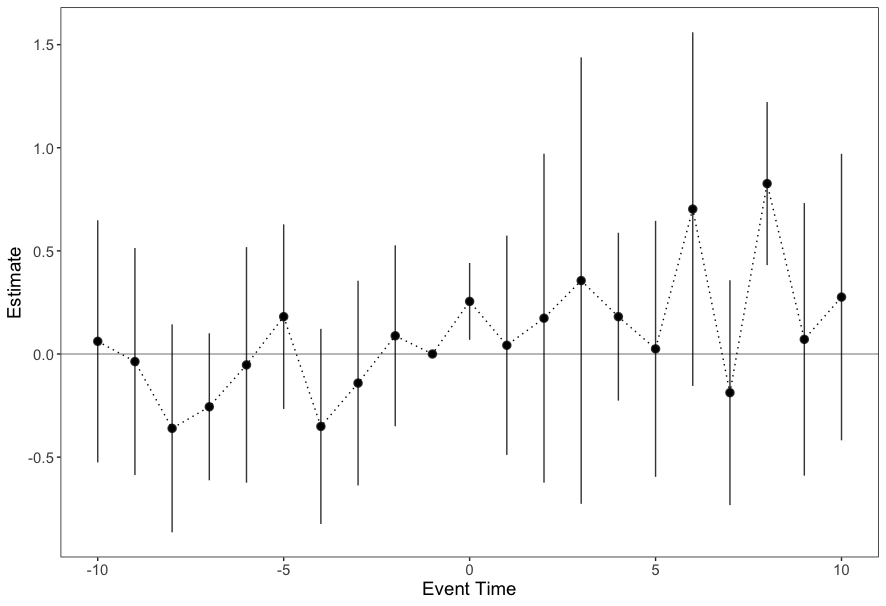}
\caption{\footnotesize{By Quarter}} 
\end{subfigure}
\smallskip
\begin{subfigure}{0.8\textwidth}
\centering
\includegraphics[width=0.9\linewidth]{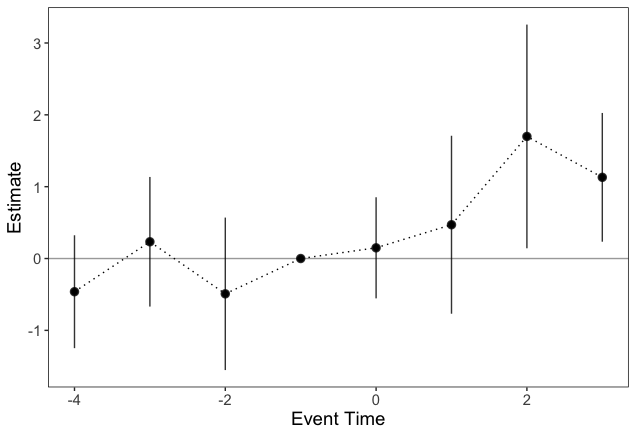}
\caption{\footnotesize{By Year}} 
\end{subfigure} 
\caption{Dynamic Effects of BlueLA Station on EV Adoption} \label{csdid_dynamic}
\footnotesize{\setstretch{0.5} \begin{flushleft} Note: This figure reports the dynamic effects of BlueLA station on EV adoption using the CSDID estimator proposed in \cite{callaway2021difference}. Dynamic effects are obtained by simple averages across groups at each period. Group is defined as the group of zipcodes that share the same starting date of its first BlueLA station.  \end{flushleft}}
\end{figure}

\newpage
\begin{figure}[H]
\centering
\includegraphics[width=0.8\linewidth]{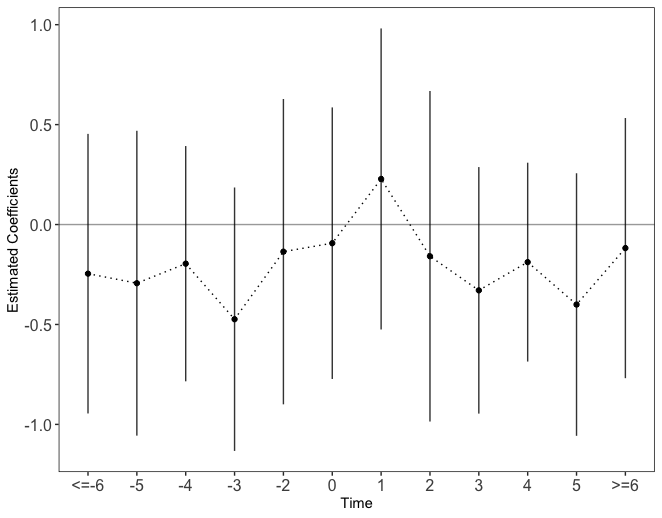}
\caption{Estimated Effects of BlueLA Station Before and After Establishment in Adjacent Zipcodes} \label{fig:coef_plot_adj}
\footnotesize{\setstretch{0.5} \begin{flushleft} Note: Standard errors are two-way cluster standard errors at year-quarter and zipcode level at 5\% significance level. \end{flushleft}}
\end{figure}

\end{document}